\documentstyle[preprint,aps]{revtex}
\begin{document}
\preprint{APCTP-1999013 // KIAS-P990912}
\def\hatt{{\hat t}}
\def\hatx{{\hat x}}
\title{Two Dimensional Anti-de Sitter Space and \\Discrete 
Light Cone Quantization}
\author{Jin-Ho Cho$^{1,2}$
\thanks{jhcho@galatica.snu.ac.kr},
Taejin Lee$^{3,4}$
\thanks{taejin@cc.kangwon.ac.kr}
and Gordon Semenoff$^{5}$
\thanks{semenoff@alf.nbi.dk}}
\address{{\it ${}^{1}$Department of Physics,
Kyung Hee University, Seoul 130-701, Korea \\
${}^2$Asia Pacific Institute for Theoretical Physics, Seoul 130-012, Korea \\
 ${}^{3}$Department of Physics, Kangwon National University, 
 Chuncheon 200-701, Korea\\
 ${}^{4}$School of Physics, Korea Institute for Advanced Study, Seoul 130-012, Korea \\
 ${}^{5}$Department of Physics, University of British Columbia, Canada 
 }}
\maketitle
\begin{abstract}
We realize the two dimensional anti-de Sitter ($AdS_2$) space as 
a Kaluza-Klein reduction of the
$AdS_3$ space in the framework of the discrete light cone quantization (DLCQ).
Introducing DLCQ coordinates which interpolate the original (unboosted) 
coordinates and the light cone coordinates, we discuss that $AdS_2/CFT$
correspondence can be deduced from the $AdS_3/CFT$.
In particular, we elaborate on the deformation of WZW model 
to obtain the boundary theory for 
the $AdS_2$ black hole. This enables us to derive the entropy of 
the $AdS_2$ black hole from that of the $AdS_3$ black hole.
\end{abstract}


\narrowtext


One of the main progresses achieved recently in the string theory is the 
$AdS/CFT$ duality \cite{mal97,adscft}, which connects the gravity in the 
$D$-dimensional anti-de Sitter ($AdS_D$) space and the $(D-1)$ dimensional 
conformal field theory ($CFT$) on its boundary. Among the $AdS/CFT$ dualities in 
the various dimensions are the $AdS_3/CFT$ and $AdS_2/CFT$ dualities
relevant for the black hole physics, since most of the black holes in the
string theories are known to contain either $AdS_3$ space or $AdS_2$
space in their near horizon geometries \cite{mal97,hyun}.
Thus, the $AdS/CFT$ dualities in low dimensions would play a key role in 
understanding the quantum aspects of the black holes.
However, compared with the case of the $AdS_3/CFT$ duality 
\cite{Mal98,tlee98,Giveon} 
the $AdS_2/CFT$ duality is less well discussed in the literature.
Observing that the near horizon geometry of the three dimensional BTZ
(Ba\~nados-Teitelboim-Zanelli) black hole \cite{btz} becomes 
effectively $AdS_2$ in the low energy
regime, one may attempt to derive the $AdS_2/CFT$ duality from the
known $AdS_3/CFT$. This approach was taken by Strominger in his 
recent work on $AdS_2/CFT$ duality \cite{Stro98}.

Here in this paper we will employ a different strategy to
derive the $AdS_2/CFT$ duality, namely the DLCQ (discrete light
cone quantization) \cite{dlcq}, which reveals the relationship between
two dualities more transparently. If the BTZ black hole is 
viewed in the light cone frame along the circle direction, the metric 
components in the
light like directions are constant and can be scaled by boosting 
the frame. Thus, if the light cone coordinate, $x^-$ is taken to be 
periodic, the Kaluza-Klein compactification can be easily performed.
In order to have a periodic light cone coordinate, we employ the DLCQ procedure,
which has been discussed recently \cite{Seib} in the context of the 
Matrix M-theory \cite{Suss}. It is found useful to introduce DLCQ coordinates, 
which interpolate the original (unboosted) coordinates and 
the light cone ones when we apply the DLCQ procedure to the
$AdS_3$ black hole. One advantage of this approach is
that we do not need to confine ourselves to the
near horizon region.


Let us begin with the well-known
$D1$-$D5$ black hole in ten dimensions, which has its near horizon 
geometry as $M_{BTZ} \times S^3 \times T^4$ \cite{horo}
\begin{eqnarray}\label{d1d5}
{ds^2\over\alpha'}&=&{U^2 \over l^2}(-dt^2+dx_5{}^2)+
{U_0{}^2 \over l^2}(\cosh{\sigma}\; dt+\sinh{\sigma}\; dx_5)^2\nonumber\\
&+&{l^2 \over U^2-U_0{}^2}dU^2
+l^2d\Omega_3+{r_1 \over r_5}\sum_{i=6}^{9}{(dx^i)^2\over\alpha'},
\end{eqnarray}
where
$x_5\sim x_5+2\pi R_s$, $x_{6,7,8,9}\sim x_{6,7,8,9}+2\pi 
V^{{1 \over 4}}\alpha'^{{1 \over 2}}$. Hereafter we will take 
$\alpha' = 1$ for the sake of convenience.
$M_{BTZ}$ corresponds to the well known three dimensional 
BTZ black hole, which has mass and angular momentum as follows
\begin{eqnarray}\label{mj}
Ml &=& \frac{1}{8Gl}(\rho^2_++\rho^2_-)=
{R_s^2 \over 8Gl^3}U_0{}^2\cosh{2\sigma}, \\
|J| &=& 2\frac{1}{8Gl} \rho_+\rho_- =
{R_s^2 \over 8Gl^3}U_0{}^2|\sinh{2\sigma}|.
\end{eqnarray}
Here we note that the BTZ black hole becomes BPS object as
we take the limit $\sigma \rightarrow \pm \infty$ while
keeping $M$, $J$ finite by making $U_0 \rightarrow 0$; in this limit
$M \mp J/l \rightarrow 0$.

It is tempting to reduce the three dimensional black hole to a 
two dimensional $AdS$ black hole by the Kaluza-Klein (KK) reduction.
The early attempt was made in ref.\cite{Lowe}, where
the four dimensional extremal Reissner-Nordstr\"om black hole
is described as $AdS_2 \times S^2$: The BTZ black hole can be
viewed as a two dimensional dilatonic $AdS$ black hole with
a $U(1)$ charge, when we apply the KK reduction along the
spatial $S^1$ direction to the BTZ black hole.
A more elaborated description of $AdS_2$ is given by Strominger 
in his recent work \cite{Stro98}, by taking the very near horizon limit
of the extremal black hole. 

In this paper we take the KK reduction along nearly lightlike circle rather 
than along the spatial circle. We first give naive idea on this. The metric 
of the BTZ black hole
in the light cone frame becomes
\begin{eqnarray}\label{light}
ds^2&=&{U_0{}^2e^{2\sigma} \over 2l^2}dx^+{}^2+
{U_0{}^2e^{-2\sigma} \over 2l^2}dx^-{}^2\nonumber\\
&+&{2U^2-U_0{}^2 \over l^2}dx^+dx^-+{l^2 \over U^2-U_0{}^2}dU^2,
\end{eqnarray}
where $x^\pm = (x^5 \pm t)/\sqrt{2}$.
It suggests that the light cone coordinate, $x^-$ may be chosen
to be compactified, since the dilaton is constant.
Here we need to employ the DLCQ procedure in order to have
a periodic light like coordinate. The DLCQ procedure
has been recently discussed \cite{Seib} in the context of the Matrix
M-theory \cite{Suss}. So some part of the analysis to
be presented also will be useful to study the Matrix
M-theory. Let us suppose that $x^5$ is periodic, 
$x^5 \sim x^5 + 2\pi R_s$, i.e., $(x^+, x^-)
\sim(x^++ \sqrt{2}\pi R_s,x^-+\sqrt{2}\pi R_s)$. 
Then by a Lorentz boost, we have 
$({x'}^+,{x'}^-)\sim({x'}^++\sqrt{2}\pi R_se^\alpha, 
{x'}^-+\sqrt{2}\pi R_se^{-\alpha}), \quad
R_s / R=(\cosh{2\alpha})^{-\frac{1}{2}}$,
where ${x'}^\pm=e^{\pm\alpha}x^\pm$ are the boosted light cone coordinates.
In the limit of the large boosting, i.e. when the boosting parameter
$\alpha \rightarrow - \infty$, (equivalently $R_s \rightarrow 0$ with
$R$ kept finite), ${x'}^-$ becomes periodic;
$({x'}^+,{x'}^-)\sim({x'}^+,{x'}^-+ 2\pi R)$.

The metric reads in terms of the boosted coordinates as, 
\begin{eqnarray}
ds^2 &=& -{2U^2(U^2-U_0{}^2) \over U_0{}^2 l^2}
e^{2(\sigma-\alpha)}
({dx'}^+)^2+{l^2 \over U^2-U_0{}^2}dU^2 \nonumber\\
&+&{U_0{}^2e^{-2(\sigma-\alpha)} \over 2l^2} 
\left({dx'}^-+ \left(2\frac{U^2}{U_0^2} -1\right)
e^{2(\sigma-\alpha)}{dx'}^+\right)^2.
\label{boost}
\end{eqnarray}
Here we observe that the dilaton factor, i.e., the compactification radius
is constant and ${x'}^+$ plays the role of time coordinate.
In the DLCQ limit, the geometry becomes $AdS_2\times S^1$.

However in order to be more transparent on the deformation of a spatial
circle to make a nearly light like circle,
we introduce new coordinates $(\hatt,U,\hatx)$, called  DLCQ  
coordinates, which interpolate the original coordinates  
$(t,U,x^5)$  (when $\alpha=0$) and the  infinitely boosted light cone 
coordinates $({x'}^+,U,{x'}^-)$ 
(when $\alpha\rightarrow-\infty$):
$(\hatt, \hatx) \sim (\hatt, \hatx + 2\pi R_s \sqrt{\cosh{2\alpha}})$, 
\begin{eqnarray}
\hatx = 
{e^{\alpha}{x'}^++e^{-\alpha}{x'}^-\over\sqrt{e^{2\alpha}+e^{-2\alpha}}}, \qquad
\hatt =
{e^{-\alpha}{x'}^+-e^{\alpha}{x'}^-\over \sqrt{e^{2\alpha}+e^{-2\alpha}}}.
\end{eqnarray} 
In the boosted frame $\hatx$ is identified as a periodic coordinate  
and the time coordinate $\hatt$ is chosen such that
$\partial_{\hatt}$ is orthogonal to $\partial_{\hatx}$.
In the infinite boosting limit ($\alpha\rightarrow-\infty$), 
$\hatt\rightarrow {x'}^{+}=e^\alpha x^+$ and 
$\hatx\rightarrow {x'}^{-}=e^{-\alpha}x^-$.

In this DLCQ coordinates, the metric reads as
\begin{eqnarray}
ds^2&=&{U^2+U_0^2\sinh^2(\sigma'+\alpha)\over l^2\cosh{2\alpha}}
\left(d\hatx+{U_0^2\sinh{2\sigma'}
-(2U^2-U_0^2)\sinh{2\alpha}
\over 2(U^2+U_0^2\sinh^2(\sigma'+\alpha))}d\hatt\right)^2\nonumber\\
&&-{U^2(U^2-U_0^2)\cosh{2\alpha}
\over l^2(U^2+U_0^2\sinh^2(\sigma'+\alpha))}d\hatt^2
+{l^2\over U^2-U_0^2}dU^2.
\end{eqnarray}
Now it becomes clear that we should keep
$\sigma' = \sigma-\alpha$ finite in order to 
obtain the metric for the space $AdS_2 \times S^1$ in the limit, 
$\alpha \rightarrow -\infty$, 
\begin{eqnarray}\label{boost2}
ds^2 &=& -{2U^2(U^2-U_0{}^2) \over U_0{}^2 l^2}
e^{2\sigma'}d\hatt^2+{l^2 \over U^2-U_0{}^2}dU^2\nonumber\\
&+&{U_0{}^2e^{-2\sigma'} \over 2l^2} \left(d\hatx+ a_\hatt d\hatt\right)^2,
\end{eqnarray}
where $a_\hatt = \left(2\frac{U^2}{U_0^2} -1\right)e^{2\sigma'}$. 
It depicts an $AdS_2$ black hole with $U(1)$ charge and
coincides with the metric Eq.(\ref{boost}) obtained by
the naive DLCQ procedure.
The mass and angular momentum of the black hole are given as
\begin{eqnarray}\label{mj2}
Ml = -J = \frac{R^2 U_0^2 e^{-2\sigma'}}{8Gl^3}.
\end{eqnarray}
Thus, the $AdS_2$ black hole can be described in terms of the
extremal $AdS_3$ black hole in the framework of DLCQ.
In passing we note that the way the DLCQ procedure results in the 
extremal limit is different from the usual one;
$\sigma \rightarrow -\infty$, and $R_s \rightarrow 0$
so that $R\,e^{-\sigma^\prime} =
R_s\, \sqrt{\cosh 2\alpha}\, e^{-\sigma+\alpha}$ is kept finite.


The DLCQ coordinates are more useful when we discuss the
boundary conformal field theory. 
In ref.\cite{tlee98}, one of the authors explicitly showed that the boundary
conformal field theory is given by a $SL(2,R)\otimes SL(2,R)$ WZW model, 
which is equivalent to the three dimensional gravity on $AdS_3$, 
resorting to the Faddeev-Shatashvili procedure \cite{Faddeev}.
Since $AdS_2$ space is obtained by the Kaluza-Klein reduction from
$AdS_3$, the same procedure would lead us to the boundary conformal
theory corresponding to the $AdS_2$ space.
We first give the
DLCQ reduction of the bulk Chern-Simons action \cite{Achu86}, which may 
be rewritten as 
\begin{eqnarray}\label{cs}
I_{CS}&=&{k \over 4\pi}\int_{ M}{{\rm tr}\epsilon^{\mu\nu}
(A_{\hatx}F_{\mu\nu}-A_\mu\partial_{\hatx}A_\nu)}
+{k \over 4\pi}\int_{\partial M}{{\rm tr} A_{\hatt}A_{\hatx}}\nonumber\\
&-&{k \over 4\pi}\int_{ M}{{\rm tr}\epsilon^{\mu\nu}(\bar{A}_{\hatx}
\bar{F}_{\mu\nu}-\bar{A}_\mu\partial_{\hatx}\bar{A}_\nu)}
-{k \over 4\pi}\int_{\partial M}{{\rm tr} \bar{A}_{\hatt}\bar{A}_{\hatx}},\nonumber
\end{eqnarray}
where $A =(\omega+{e \over l})_I{}^Adx^IJ_A$,
$\bar{A}=(\omega-{e \over l})_I{}^Adx^I\bar{J}_A$, 
$\left[J_A, J_B\right] = \epsilon_{AB}{}^{C} J_C$,
$\left[\bar{J}_A, \bar{J}_B\right] = \epsilon_{AB}{}^{C} \bar{J}_C$,
$\mu,\nu,\rho\in\{\hatt,U\}$,
$A,B,C\in\{0,1,2\}$, and $I,J,K\in\{\hatt,U, \hatx^5\}$.
We may cast the dreibein and
the spin connection into the form of Kaluza-Klein ansatz as
\begin{eqnarray}
e_I{}^A=\pmatrix{
e_\mu{}^a & -le^{\psi}a_\mu \cr
0 & -{l\over R}e^\psi \cr
},\qquad\omega_I{}^A=\pmatrix{
\omega_\mu{}^a & \omega_\mu{}^2 \cr
\omega_{\hatx}{}^a & \omega_{\hatx}{}^2 \cr
},
\nonumber
\end{eqnarray}
where $a\in \{0,1\}$. Then, imposing the torsion free conditions, which are obtained
as part of the equations of motion
we find that the Chern-Simons action 
reduces to the action for the two dimensional gravity as expected
\begin{eqnarray}\label{2dg}
I_{2D}={k \over 2}\int\sqrt{-g}\left(e^\psi({ R}+{2\over l^2})-
{e^{3\psi}l^2 \over 4}f_{\mu\nu}f^{\mu\nu}
\right),
\end{eqnarray}
where $f_{\mu\nu}=\partial_\mu a_\nu-\partial_\nu a_\mu$.


From the metric for the BTZ black hole in the light cone frame Eq.(\ref{light}), 
we obtain the black hole solution in terms of the gauge fields as 
follows.
\begin{eqnarray}
A^0 &=& 0, \quad A^1 = 0, \nonumber\\
A^2 &=& {U_0e^{-\sigma'}\over l^2}
{-e^{\alpha}d\hatt+e^{-\alpha}d\hatx
\over\sqrt{\cosh{2\alpha}}},\\
\bar{A}^0 &=& -2{Ue^{\sigma'}\over U_0l^2}
\left(U^2-U_0^2\right)^{1\over 2}
{e^{-\alpha}d\hatt+e^{\alpha}d\hatx
\over\sqrt{\cosh{2\alpha}}},\nonumber\\
\bar{A}^1 &=& - 2 \left(U^2 - U_0{}^2\right)^{-{1\over 2}}dU \nonumber\\
\bar{A}^2 &=& -2{e^{\sigma'}\over U_0l^2}\left(U^2-{U_0^2\over 2}\right)
{e^{-\alpha}d\hatt+e^{\alpha}d\hatx
\over\sqrt{\cosh{2\alpha}}}.\nonumber
\end{eqnarray}
Among those, only the components $A_\hatx$ and $\bar{A}_\hatx$ are 
physically important because the nontrivial geometrical structure of 
the three dimensional space is completely encoded
by holonomies or Wilson loops of the Chern-Simons gauge 
fields \cite{holo},
$W[C] = {\cal P} \exp \left( \oint_C A_\hatx d\hatx \right), \quad
{\bar W}[C] = {\cal P} \exp \left( \oint_C {\bar A}_\hatx d\hatx \right)$,
where $C$ is a closed curve and ${\cal P}$ denotes a path ordered product.
Since in the limit where $\alpha \rightarrow -\infty$,
$ A_\hatx \rightarrow \sqrt{2}{U_0e^{-\sigma'}\over l^2},\quad
{\bar A}_\hatx \rightarrow 0$,
we see that the the right $SL(2,R)$ sector becomes trivial while
the left $SL(2,R)$ sector remains relevant.
It is consistent with the observation that the
DLCQ and scaling procedure results in the BPS limit.
This also explains how the isometry group of $AdS_3$, $SL(2,R)
\otimes SL(2,R)$ reduces to the isometry group of $AdS_2$, $SL(2,R)$
in the framework of DLCQ.

Rewriting the boundary action obtained in ref.\cite{tlee98} in terms of 
the DLCQ coordinates, we find that the boundary theory for $AdS_2$ may 
be given by the WZW model, deformed by the DLCQ procedure. The total action is 
composed of the followings;
\begin{eqnarray}\label{baction}
I &=& {k\over 4\pi}\int_{M}{\rm tr}\epsilon^{ij}\partial_{\hatt}A_iA_j 
+{k\over 4\pi}\int_{M}{\rm tr}\epsilon^{ij}A_{\hatt}F_{ij}
\nonumber\\  
&-&{k\over 4\pi}\int_{\partial M}
{\rm tr}\left[e^{2\alpha}(A_{\hatx}+\partial_{\hatx}gg^{-1})^2
+2A_{\hatt}(A_{\hatx}+\partial_{\hatx}gg^{-1})\right] \nonumber\\
&-&{k\over 4\pi}\int_{\partial M}
{\rm tr}\partial_{\hatt}gg^{-1}\partial_{\hatx}gg^{-1}
-{k \over 12\pi} \int_{M}{\rm tr}(g^{-1}dg)^3,\nonumber\\
{\bar I} &=& -{k\over 4\pi}\int_{M}
{\rm tr}\epsilon^{ij}\partial_{\hatt}\bar{A}_i\bar{A}_j
-{k\over 4\pi}\int_{M}{\rm tr}\epsilon^{ij}\bar{A}_{\hatt}\bar{F}_{ij}\nonumber\\
&-&{k\over 4\pi}\int_{\partial M}
{\rm tr}\left[e^{-2\alpha}(\bar{A}_{\hatx}+\partial_{\hatx}\bar{g}\bar{g}^{-1})^2
-2\bar{A}_{\hatt}(\bar{A}_{\hatx}
+\partial_{\hatx}\bar{g}\bar{g}^{-1})\right]
\nonumber\\
&+&{k\over 4\pi}\int_{\partial M}
{\rm tr}\partial_{\hatt}\bar{g}\bar{g}^{-1}
\partial_{\hatx}\bar{g}\bar{g}^{-1}
+{k \over 12\pi}\int_{M}{\rm tr}(\bar{g}^{-1}d\bar{g})^3,
\end{eqnarray}
where $i,j\in (U,\hatx)$.
With $(\tau, \theta)=\frac{1}{R}(e^{2\alpha}\hatt, \hatx)$
we may identify $I$ as the left sector of the action for the
BTZ black hole given in ref.\cite{tlee98}.
We also find that ${\bar I}$ can be understood as the right 
sector of the action for the BTZ black hole in terms of
$(\tau', \theta)=\frac{1}{R}(e^{-2\alpha}\hatt, \hatx)$. 
The right sector is confined in the extreme low energy regime,
thus, suppressed. The left $SL(2,R)$ sector only becomes relevant 
to the $AdS_2$.

The DLCQ procedure presented here illustrates explicitly how $AdS_2/CFT$ 
correspondence can be deduced from the $AdS_3/CFT$. It would be
interesting to explore its consequences in various contexts.
One may find an application of the DLCQ procedure readily in 
evaluation of the entropy for the $AdS_2$ black hole. 
To evaluate the entropy, we adopt the result of
ref.\cite{tlee98}, where the entropy of the three dimensional
BTZ black hole is evaluated in accord with the proposal of
Strominger \cite{stro}. Comparing the boundary action for the
$AdS_2$ (\ref{baction}) with that for the BTZ black hole, given
in ref.\cite{tlee98}, we find that the
the expectation values of the Virasoro generators $L_0$ and 
${\bar L}_0$ for the $AdS_2$ black hole are given by 
\begin{eqnarray}
&&n_L=<L_0> = \frac{k{U_0}^2 R^2 e^{-2\sigma '}}{2l^4} ,\nonumber\\
&&n_R=<{\bar L}_0> =  
\frac{k{U_0}^2 R^2 e^{2\sigma '}}{2l^4}e^{4\alpha}.
\end{eqnarray}
Here we assume that $n_L = n_R= 0$ for an appropriately chosen
vacuum state of the black hole.
Then it follows from the Cardy formula that the entropy for the 
$AdS_2$ black hole is evaluated as
\begin{eqnarray}\label{ent}
S 
= \sqrt{2}\pi k \frac{RU_0 e^{-\sigma'}}{l^2} +
\sqrt{2}\pi k \frac{RU_0 e^{\sigma'}}{l^2} e^{2\alpha}.
\end{eqnarray} 
As $\alpha \rightarrow -\infty$,
the right sector does not contributes to the entropy, 
$S \rightarrow 4\pi k \sqrt{GM}$.
(Some difficulty in evaluating the entropy by
applying the Cardy formula to the boundary conformal theory
was pointed out by Carlip \cite{carlip}. 
This difficulty may be resolved if we consider the space-time 
Virasoro algebra, regarding the boundary conformal theory as
the worldsheet action for the string on $AdS_3$ \cite{Giveon}.
It would be worth while to extend the work of ref.\cite{tlee98}
along this direction.)

We conclude this paper with a few brief remarks. We show that the 
$AdS_2$ black hole can be described as a DLCQ limit of the $AdS_3$
black hole. The boundary theory for $AdS_2$ is found to be 
a $SL(2,R)$ WZW model defined on the light like coordinate.
If we are concerned with the gravity on $AdS_2$,
the zero mode sector of the $AdS_3$ suffices in the
framework of DLCQ. In accordance with it, the zero mode sector of 
the WZW model given by the action Eq.(\ref{baction}) defines
the boundary theory, hence a $SL(2,R)$ quantum mechanics.
The correspondence between the gravity on $AdS_2$ and the
$SL(2,R)$ boundary conformal quantum mechanics will be discussed
in detail elsewhere. We note that the entropy of the $AdS_2$ black hole
evaluated as Eq.(\ref{ent}) is the same as that of the $AdS_3$ black hole
in the original frame. Thus, the entropy is preserved by the DLCQ procedure.

The present paper may be extended along various directions.
One of the direction would be the DLCQ reduction of the string on 
$AdS_3$ \cite{Giveon}. It would be interesting to see if the DLCQ reduction 
leads us to a point source on $AdS_2$ and the entropy of $AdS_2$ 
black hole can be given by the quantum mechanics. This may 
clarify some subtle issues associated with the entropy of black holes 
in $AdS_3$ and $AdS_2$. 
The two dimensional black holes have been one 
of the important subjects in string theory and gravity.
The present work may enable us to
discuss the various aspects of the two dimensional black holes 
\cite{2dbh1,2dbh2} from the viewpoint of the three dimensional black hole.
According to the conjecture of $AdS/CFT$ correspondence \cite{mal97}
the (2+1) dimensional gravity on $AdS_3$ is supposed to be equivalent to the
boundary (1+1) dimensional conformal Yang-Mills theory on its boundary.
The present work also suggests that the boundary conformal theory corresponding
to the gravity on $AdS_2$ may be obtained by the DLCQ reduction of the 
(1+1) dimensional Yang-Mills theory. Work along this direction
is now in progress \cite{chl99}.
The most important
application of the present work may be found in the Matrix M-theory.
DLCQ reduction of the M-brane configurations may shed some light on the
Matrix M-theory.
After completing the work, we found that reduction of $AdS_3$ to
$AdS_2$ along the light like coordinate also has been suggested in the
study of M-brane configurations \cite{sken}.

\section*{Acknowledgement}
The work of TL was supported in part by the Basic Science Research 
Institute Program, Ministry of Education of Korea (BSRI-98-2401). 
Part of the work of TL was done during his visit to PIMS, KEK and YITP.
We would like to thank K. Skenderis for informing us of ref.\cite{sken}
and useful comments.

\end{document}